\documentstyle[prl,aps,epsf,epsfig]{revtex}
\twocolumn
\begin{document}

\title{Polarization Observables in Hard Rescattering Mechanism 
of Deuteron Photodisintegration} 
\author{Misak~M.~Sargsian}

\address{Florida International University, Miami, FL 33199 USA}

\date{\today}

\maketitle

\begin{abstract}
{\bf Abstract:} Polarization properties of high energy photodisintegration 
of the deuteron are studied within the framework of the hard rescattering 
mechanism~(HRM). In HRM, a  quark of one nucleon knocked-out by 
the incoming photon rescatters with a quark of the other nucleon leading 
to the production of two nucleons with high relative momentum.  
Summation of all relevant quark rescattering amplitudes allows us to 
express the scattering amplitude of the reaction through the convolution 
of a hard photon-quark interaction vertex, the large angle p-n scattering amplitude 
and the low momentum deuteron wave function.  Within HRM, it is demonstrated that 
the polarization observables in hard photodisintegration of the deuteron can be 
expressed through the five helicity amplitudes of NN scattering at high momentum 
transfer. At 90$^\circ$ CM scattering HRM predicts the dominance of the isovector channel of 
hard $pn$ rescattering, and it explains the observed smallness of induced, $P_y$ and 
transfered, $C_x$  polarizations without invoking the argument of helicity conservation. 
Namely, HRM predicts that $P_y$ and $C_x$ are proportional to the $\phi_5$ helicity 
amplitude which vanishes at $\theta_{cm}=90^\circ$ due to symmetry reasons.  
HRM predicts also a nonzero value for $C_z$ in the 
helicity-conserving regime and a positive $\Sigma$ asymmetry which is related 
to the dominance of the isovector channel in the hard reinteraction. We extend our 
calculations to the region where large polarization effects are observed in $pp$ 
scattering as well as give predictions  for angular dependences. 

\end{abstract}

\bibliographystyle{unsrt}

\noindent
\section{Introduction}
Hard photodisintegration of the deuteron provides a unique tool for studying 
the role of quarks and gluons in  nuclear interactions. During the last 
decade several experiments have been performed\cite{NE8,NE17,E89012,angular,polar,Yerevan}
which indicated strongly the importance of quark-gluon degrees of freedom  
in these reactions starting at $E_{\gamma}\ge 1$~GeV.

First QCD based predictions for high momentum transfer photodisintegration of the 
deuteron were done within minimal Fock component approximation\cite{BF,MMT} in which 
it is  assumed that only minimal number of partonic constituents dominate in large angle 
hard two-body scattering. Within this approximation the energy dependences of the set of 
fixed angle hard two-body reactions can be predicted according to the counting rule:
${d\sigma\over dt}\sim s^{-(n_1+n_2+n_3+n_4-2)}$, in which $n_i$ is the number of 
fundamental constituents in the particle $i$  which is involved in the reaction. 
This prediction has been  confirmed  experimentally practically for all 
two-body reactions  for fixed angle hard scattering kinematics in 
which $-t,-u \ge 2$~GeV$^2$.

For high momentum transfer $\gamma+d\rightarrow p+n$ reaction the above counting rule 
predicts an  energy dependence, $\sim s^{-11}$~\cite{BCh}, which was confirmed 
experimentally for photon energies starting at $1$~GeV\cite{NE8,NE17,E89012}.

The minimal Fock component approximation can be proven rigorously within 
perturbative QCD (pQCD) in which the masses of interacting current 
quarks are neglected.  Thus the 
experimental success of the minimal Fock component approximation raised the expectations 
that the observed energy dependences indicate  the onset of pQCD regime. 
This was an important question since there were several arguments\cite{IsgLS,Rad} 
against the application of pQCD in the considered energy range  as well as the attempts 
to describe the absolute cross sections of hard two-body exclusive reactions within 
leading twist pQCD 
have been largely unsuccessful (see e.g.\cite{Farrar,BDixon}) underestimating the observed 
cross sections by several orders of magnitude \footnote{The smallness of the calculated 
cross sections does not rule out completely the relevance of pQCD regime, since one may 
expect a sizable effects from unaccounted hidden color component of hadronic wave 
functions\cite{Brodsky}.}.

Since, in QCD the interaction is realized through the exchange of vector gluons, 
in pQCD (due to vanishing quark masses) the helicity of interacting particles should 
be conserved. Therefore as an independent check of the onset of pQCD one can 
investigate the effects of  hadronic helicity conservation~(HHC).

The experiments  which are aimed at the studies of polarization observables in hard  
reactions are  best suited for HHC studies. The first experiments were performed for 
elastic $pp$ scattering. While in wide range of hard scattering kinematics the $pp$ 
data generally are in agreement with HHC, in some instances the striking disagreement 
is observed\cite{Crabb}. For example in ${\vec p}+{\vec p}\rightarrow p+p$ scattering  
at $\theta_{cm}=90^\circ$ and   $P_{Lab}=11.75$~GeV\cite{Crabb} the measurements 
demonstrated that protons polarized transverse to the scattering plane  have 
four times larger probability to scatter with spins parallel than antiparallel 
to each other. This number is considerably larger than HHC predication 
of two\cite{FGST,BCL}. Several theoretical approaches have been proposed 
to describe the observed enhancement of the polarization effects
(see e.g.\cite{FGST,BCL,BrodTer,CChM,RS}) however 
the experimental evidence is very limited for  meaningful progress in 
understanding the mechanism of  HHC violation.

Since the onset of energy scaling in the cross section of  deuteron photodisintegration  
is observed already at $E_\gamma\ge 1$~GeV and $\theta_{cm}=90^\circ$, the measurement of 
polarization observables at the same kinematics  will suit ideally for HHC studies.
There were several recent studies \cite{polar,Yerevan,GG,GKCDSMRR} in 
polarization properties of  high energy deuteron photodisintegration. With JLAB building 
up a systematic experimental program on  deuteron photodisintegration with polarization  
measurements one may expect a wealth of the new data within  next several years\cite{polar,RG}. 

In this paper  we study several polarization observables in hard photodisintegration 
reaction of the deuteron within the recently developed model of hard rescattering 
(HRM)\cite{gdpn}. HRM is based on the assumption that hard photodisintegration of the 
deuteron proceeds through two steps: at first, the incoming photon knocks-out a quark from 
one nucleon in the deuteron which then makes a hard rescattering with a quark of the second 
nucleon in the deuteron. This assumption allows us to express the disintegration amplitude 
through the convolution of the deuteron wave function, hard photon-quark interaction 
amplitude  and the amplitude of hard $pn$ scattering.  The latter was estimated using 
the experimental $pn$ scattering  data. HRM provides also a convenient framework 
for calculation of  the polarization observables of photodisintegration reaction, 
expressing them through the helicity amplitudes of $pn$ scattering. In the next sections 
within HRM we calculate several polarization observables which are currently investigated 
experimentally. HRM gives rather different insight on observed regularities in polarization 
measurements and makes several predictions whose verification can advance our 
understanding the dynamics of hard photodisintegration.

\section{Hard Rescattering Mechanism}

We are considering a reaction
\begin{equation}
\gamma + d \rightarrow p + n
\label{reaction}
\end{equation}
in which the {\em polarizations} of $\gamma$ and/or $p$ are measured. 
The hard scattering is defined by a requirement that $-t,-u\ge 2$~GeV$^2$, where 
$t=(q-p_p)^2=(p_n-p_d)^2$, $u = (q-p_n)^2  = (p_p-p_d)^2$ and $q$, $p_d$, $p_p$ and $p_n$ 
are four--momenta of incoming photon, target deuteron, outgoing proton and neutron  
respectively.

Within HRM\cite{gdpn} it is assumed that final two high--$p_T$ nucleons are produced due 
to hard rescattering of a quark, knocked out by incoming photon 
from one nucleon, with a quark in other nucleon. As a result the  sum of diagrams similar 
to the one presented in Fig.1  gives the  main contribution to the scattering amplitude of 
the reaction (\ref{reaction}).

\begin{figure}[ht]
\vspace{-0.6cm}
\centerline{\epsfig{height=4cm,width=9cm,file=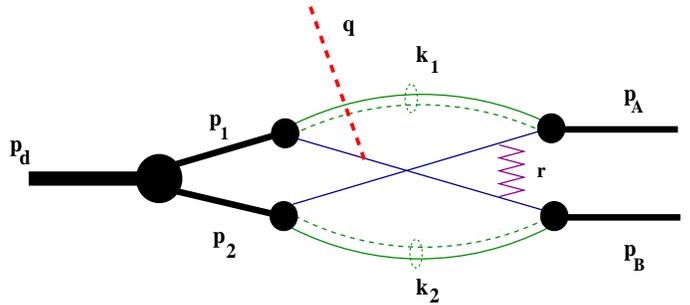}}
\caption{Typical quark-rescattering diagram}
\label{Fig.1}
\end{figure}
 We start with analyzing the scattering amplitude corresponding to 
the diagram of Fig.\ref{Fig.1}: 
\begin{eqnarray}
& & \langle\lambda_A,\lambda_B\mid A \mid \lambda_\gamma, \lambda_D\rangle = 
\sum\limits_{(\eta_1,\eta_2),(\xi_1,\xi_2),(\lambda_1,\lambda_2)\zeta} \int 
\nonumber \\ & & 
\left\{ {\psi^{\dagger\lambda_B,\eta_2}_N(p_B,x'_2,k_{2\perp})\over 1-x'_2}\bar 
u_{\eta_2}(p_B-k_2) [-igT^F_c\gamma^\nu]\times\right. \nonumber \\ & & 
{i[u_\zeta(p_1-k_1+q)\bar u_\zeta(p_1-k_1+q)\over (1-x_1)s'(\alpha_c-\alpha+i\varepsilon)}
[-ie_q{\bf \epsilon^{\lambda_\gamma}_\mu\gamma^\mu}]u_{\xi_1}(p_1-k_1)\nonumber \\ & & 
\left. {\psi_N^{\lambda_1,\xi_1}(p_1,x_1,k_{1\perp})\over (1-x_1)}\right\}_1\left\{ 
{\psi^{\dagger\lambda_A,\eta_1}_N(p_A,x'_1,k_{1\perp})\over 1-x'_1}\right.  
\nonumber \\ & & 
\left. \bar u_{\eta_1}(p_A-k_1)[-igT^F_c\gamma^\mu]u_{\xi_2}(p_2-k_2)
{\psi_N^{\lambda_2,\xi_2}(p_2,x_2,k_2)\over (1-x_2)} \right\}_2 \nonumber \\ & & 
G^{\mu,\nu}(r) {dx_1\over x_1}{d^2k_{1\perp}\over 2 (2\pi)^3}
{dx_2\over x_2}{d^2k_{2\perp}\over 2 (2\pi)^3}
{\Psi_D^{\lambda_D,\lambda_1,\lambda_2}(\alpha,p_\perp)\over (1-\alpha)}{d\alpha\over 
\alpha}
{d^2p_\perp\over 2(2\pi)^3} \nonumber \\
\label{ampl0}
\end{eqnarray}
were the four-momenta: $p_1$, $p_2$, $k_1$, $k_2$, $r$, $p_A$ and $p_B$ 
are defined in Fig.\ref{Fig.1}. Note that $k_1$ and $k_2$ define
the four-momenta of residual quark-gluon system of the nucleons without specifying 
their actual composition. $s' = s-M_d^2$, where $s = (q+p_d)^2$.  $x_1$, $x'_1$, 
$x_2$ and $x'_2$ are the light-cone  momentum fractions of initial and final nucleons 
carried out by spectator system in the nucleons ($x_{1(2)} = {k_{1(2)+}\over p_{1(2)+}}$, 
$x'_{1(2)} = {k_{1(2)+}\over p_{A(B)+}}$)\footnote{The light cone four-momentum is 
defined as $(p_+,p_-,p_\perp)$, where $p_{\pm} = E\pm p_z$. Here the $z$ axis is defined 
in the direction opposite to the incoming photon momentum.}. 
$\alpha={p_{2+}\over p_{d+}}$ is   
the light cone momentum fraction of the deuteron carried by one of the nucleons and 
$p_\perp$ is the relative transverse momentum of the nucleons in the 
deuteron.
The denominator $(1-x_1)s'(\alpha_c-\alpha+i\varepsilon)$ is obtained from 
the denominator of knocked-out quark propagator, $(p_1-k_1+q)^2-m_q^2+i\varepsilon$ 
by expressing it through $\alpha$ and 
\begin{equation}
\alpha_c = 1 + {1\over s'}\left[\tilde m_N^2 - 
{\tilde m_R^2(1-x_1)+m_q^2x+(k_1-xp_1)^2\over x_1(1-x_1)}\right]
\label{alpc}
\end{equation}
where $\tilde m_N^2 = p_{1-}p_{d+}(1-\alpha)- p^2_\perp$ and 
$\tilde m_R^2 = k_{1-}(p_{d+}(1-\alpha)x_1-k_{1\perp}^2$ are an effective masses of 
the off-shell nucleon and its residual system respectively. $m_q$ represents the current 
quark mass of the knocked-out quark. 
The scattering process in Eq.(\ref{ampl0}) can be described through the combination 
of the following blocks: a)~$\Psi_D^{\lambda_D,\lambda_1,\lambda_2}(\alpha,p_\perp)$, 
is the light-cone deuteron wave function which describes the transition of the deuteron with 
helicity $\lambda_D$ into two nucleons with $\lambda_1$ and $\lambda_2$ helicities 
respectively. b)~ The term in $\{....\}_{1}$ describes the ``knocking out'' a  
 $\xi_1$-helicity quark  from the $\lambda_1$-helicity nucleon by an incoming 
photon with helicity $\lambda_\gamma$.
Subsequently, the ``knocked-out''  $\zeta_1$-helicity quark exchanges gluon,  
($[-igT^F_c\gamma^\nu]$), with a quark from second nucleon producing a final 
$\eta_2$-helicity quark which enters the nucleon B with helicity $\lambda_B$. 
c)~The term in $\{...\}_{2}$ describes the emerging $\xi_2$-helicity
quark from $\lambda_2$ - helicity nucleon
which  then exchanges a gluon, ($[-igT^F_c\gamma^\mu]$), with the knocked-out 
quark and produces a final $\eta_1$-helicity quark which enters the  nucleon with helicity 
$\lambda_A$.  d)~The propagator of the exchanged gluon is 
$G^{\mu\nu}(r) = {d^{\mu\nu}\over r^2+i\varepsilon}$ with polarization matrix, 
$d^{\mu\nu}$, (fixed by light-cone gauge), and $r=(p_2-k_2+l)-(p_1-k_1+q)$, with 
$l = (p_B-p_2)$. In Eq.(\ref{ampl0}) the $\psi^{\lambda,\tau}_N$ represents everywhere a 
$\tau$-helicity  single quark  wave function of $\lambda$-helicity nucleon and 
$u_\tau$ is the quark spinor defined in the helicity basis.
We keep only the $u_\zeta \bar u_\zeta$ term in the numerator of the 
knocked-out quark propagator, since this is the only term that contributes 
through the soft~(dominant) component of the deuteron wave function.

Next we integrate Eq.(\ref{ampl0}) by $\alpha$, taking into account only on-mass 
shell contribution of struck quark propagator, i.e. the second term in the decomposition:
$(\alpha_c-\alpha+i\varepsilon)^{-1} = 
{\cal P}(\alpha_c-\alpha)^{-1} - i\pi\delta(\alpha_c-\alpha)$. The on-mass shell 
approximation allows us to evaluate the photon-quark interaction vertex, for which, 
in vanishing current quark mass approximation one obtains:
\begin{eqnarray}
& & \bar u_\zeta(p_1-k_1+q)[-ie_q{\bf \epsilon^{\lambda_\gamma}_\mu\gamma^\mu}]
u_{\xi_1}(p_1-k_1)
 = e_q \sqrt{2s'}\times
\nonumber\\
& & \sqrt{\left[1-(1-\alpha)(1-x_1)\right](1-\alpha)(1-x_1)}
\times \delta^{\zeta,\lambda_{\gamma}}\delta^{\lambda_{\gamma},\xi_1}.
\nonumber\\
\label{hcon}
\end{eqnarray}
Two important features of the above equation should be emphasized: i) 
an energetic photon selects only those quarks from a nucleon that have the  
same helicity that the photon has ($\xi_1=\lambda_\gamma$); ii)the  helicity of 
the initial quark is conserved after it was struck by incoming photon 
$(\zeta = \xi_1)$. Inserting Eq.(\ref{hcon}) into Eq.(\ref{ampl0}) and taking 
the $d\alpha$ integral by estimating it through the residue at the 
pole $\alpha=\alpha_c$ one obtains:
\begin{eqnarray}
& & \langle \lambda_A,\lambda_B\mid  A\mid \lambda_\gamma, \lambda_D\rangle = 
\sum\limits_{(\eta_1,\eta_2),(\xi_2),(\lambda_1,\lambda_2)} \int 
{e_q \sqrt{2}\over (1-x_1)\sqrt{s'}}\nonumber \\ 
& &  \sqrt{[1-(1-\alpha_c)(1-x_1)](1-\alpha_c)(1-x_1)}
\nonumber \\ & & 
\left\{ {\psi^{\dagger\lambda_B,\eta_2}_N(p_B,x'_2,k_{2\perp})\over 1-x'_2}\bar 
u_{\eta_2}(p_B-k_2) [-igT^F_c\gamma^\nu]\cdot\right.
\nonumber \\ & & 
u_{\lambda_\gamma}(p_1-k_1+q)
{\psi_N^{\lambda_1,\lambda_\gamma}(p_1,x_1,k_{1\perp})\over (1-x_1)}\times
\nonumber \\ & &  
{\psi^{\dagger\lambda_A,\eta_1}_N(p_B,x'_1,k_{1\perp})\over 1-x'_1}
\bar u_{\eta_1}(p_A-k_1)[-igT^F_c\gamma^\mu]u_{\xi_2}(p_2-k_2) 
\nonumber \\ &&
\left.
{\psi_N^{\lambda_2,\xi_2}(p_2,x_2,k_2)\over (1-x_2)} 
G^{\mu,\nu}(r) {dx_1\over x_1}{d^2k_{1\perp}\over 2 (2\pi)^3}
{dx_2\over x_2}{d^2k_{2\perp}\over 2 (2\pi)^3}\right\}
\nonumber \\ &  &
{\Psi^{\lambda_D,\lambda_1,\lambda_2}(\alpha,p_\perp)\over (1-\alpha)\alpha}
{d^2p_\perp\over 4(2\pi)^2}. 
\label{ampla}
\end{eqnarray}
One can relate 
the expression in $\{....\}$ to the quark-interchange kernel of $NN$ interaction. 
Taking into account the fact that the deuteron wave  function peaks strongly at 
$\alpha_c={1\over 2}$ we approximate Eq.(\ref{ampla}), choosing $\alpha_c={1\over 2}$. 
In this case in $x_1\rightarrow 0$ limit, which corresponds to the Feynman picture of 
hard scattering\cite{Feynman}, Eq(\ref{ampla}) factorizes into the product of 
$\gamma- quark$ scattering vertex and quark-exchange amplitude of $NN$ scattering\cite{gdpn}. 
In the case of the minimal Fock component approximation, in which $x_1, (1-x_{1})\sim 1$ 
the factorization is correct up to the scaling function 
$f(\theta_{cm})$, with $f(\theta_{cm}=90^\circ)\approx 1$\cite{gdpn}. 
Using this factorization, for Eq.(\ref{ampla}) one obtains:
\begin{eqnarray}
& &\langle \lambda_A,\lambda_B,\mid A_{Q_i} \mid \lambda_\gamma,\lambda_D\rangle = 
\sum\limits_{(\eta_1,\eta_2),(\xi_2),(\lambda_1,\lambda_2)} \int 
{eQ_i f({\theta_{cm}})\over \sqrt{2s'}} \nonumber \\ 
& & 
\ \ \ \ \langle \eta_2,\lambda_B|\langle \eta_1,\lambda_A|A^i_{QIM}(s,l^2)
|\lambda_1,\lambda_\gamma\rangle |\lambda_2 \xi_2\rangle \nonumber \\
& &
\ \ \ \ \Psi^{\lambda_D,\lambda_1,\lambda_2}(\alpha_c,p_\perp) {d^2p_\perp\over (2\pi)^2} 
\label{amplb}
\end{eqnarray}
where $\langle \eta_2, \lambda_B|\langle\eta_1, \lambda_A|A^i_{QIM}({s,l^2})
|\lambda_1,\lambda_\gamma\rangle|\lambda_2,\xi_2\rangle$ is 
the quark-interchange kernel (with quark-$i$ interacting with the photon)
corresponding to the expression in $\{...\}$ in 
Eq.(\ref{ampla}). Here $|\lambda, \eta\rangle$ represents 
$\eta$-helicity quark wave function of $\lambda$-helicity nucleon. Since 
the momenta of interacting quarks 
are large ($1-x_1\sim 1$) one can  assume that the interchanging 
quarks carry the helicities of a parent nucleons (i.e. $\eta = \lambda$). 
This allows us to express the scattering amplitude in 
Eq.(\ref{amplb}) through the helicities of the photon, deuteron and  nucleons 
as follows: 
\begin{eqnarray}
& &\langle \lambda_A,\lambda_B \mid A_{Q_i}\mid \lambda_\gamma,\lambda_D\rangle = 
\sum\limits_{\lambda_2} \int 
{ef({\theta_{cm}})\over \sqrt{2s'}} Q_i\times\nonumber \\ 
& & 
\langle \lambda_A,\lambda_B|A^i_{QIM}({s,l^2})|
\lambda_\gamma,\lambda_2\rangle 
\Psi^{\lambda_D,\lambda_\gamma,\lambda_2}(\alpha_c,p_\perp) {d^2p_\perp\over (2\pi)^2} 
\label{amplc}
\end{eqnarray}
where $|\lambda_1,\lambda_2\rangle$ represents two nucleons having 
$\lambda_1$ and $\lambda_2$ helicities respectively.
Note that $A^i_{QIM}$ in the above equation is weighted with the charge of 
the knocked-out quark $Q_i$, thus it can not be directly related to
the quark interchange amplitude of $pn\rightarrow pn$ scattering. 

To calculate the total scattering amplitude within HRM we sum all amplitudes of 
topologies of Fig.{\ref{Fig.1}. 
Identifying $\lambda_A$ and $\lambda_B$ with the helicities of proton and neutron 
respectively, one obtains:
\begin{eqnarray}
\langle p_{\lambda_A},n_{\lambda_B} \mid A\mid \lambda_\gamma,\lambda_D\rangle & = & 
\sum\limits_{i \in D}\left[ 
\langle p_{\lambda_A},n_{\lambda_B}\mid A_{Q_i}\mid \lambda_\gamma,\lambda_D\rangle\right. 
- \nonumber \\ 
 & & \left. \langle n_{\lambda_B},p_{\lambda_A}\mid A_{Q_i}\mid \lambda_\gamma,
\lambda_D\rangle\right]
\label{amplsum}
\end{eqnarray}
where one sums valence quarks of the deuteron. This sum can be performed within 
the quark-interchange model of hadronic interactions, which allows  
us to represent the $NN$ scattering amplitude as follows\cite{BCL}:
\begin{equation} 
\langle a'b' |A|ab\rangle = 
{1\over 2}\langle a'b'| \sum\limits_{i\in a\ , \ j\in b} 
[I_iI_j + \vec \tau_i\vec\cdot\tau_j] F_{i,j}(s,t)|ab\rangle
\label{NNQIM}
\end{equation}
where $I_i$ and  $\tau_i$ are identity and  Pauli matrices 
defined in $SU(2)$ flavor (isospin) space of the interchanged 
quarks. The kernel, $F_{i,j}(s,t)$ describes an interchange of $i$ and $j$ 
quarks\footnote{The additional assumption of helicity conservation allows us to 
express the kernel in the form\cite{BCL}: 
$F_{i,j}(s,t) = {1\over 2}[I_iI_j + \vec \sigma_i\vec\cdot\sigma_j]\tilde F_{i,j}(s,t)$, 
where $I_i$ and  $\sigma_i$ operate in $SU(2)$ helicity ($H$-spin) space of 
exchanged ($i,j$) quarks\cite{BCL}. However for our discussion the assumption of 
helicity conservation is not required.}.

One can use Eq.(\ref{NNQIM}) to calculate the quark-charge weighted QIM 
amplitude, $\langle a'b' |A^{Q}|ab\rangle$,  to obtain:
\begin{eqnarray}
& & \langle a'b' |A^Q|ab\rangle \mid_{a,b \in D}  \ = \nonumber \\
& & {1\over 2}\langle a'b'| \sum\limits_{i\in a\ , \ j\in b} 
[I_iI_j + \vec \tau_i\vec\cdot\tau_j] (Q_i+Q_j)F_{i,j}(s,t)|ab\rangle = \nonumber \\
&  & (Q_u+Q_d)\langle a'b' |A|ab\rangle  = {1\over 3}\langle a'b' |A|ab\rangle.
\label{QNNQIM}
\end{eqnarray}
The above result can be understood qualitatively: since the number of $u$ and $d$ 
quarks in the deuteron are equal one has the same number of diagrams with knocked out
$u$ and $d$ quarks. Using Eqs.(\ref{amplc},\ref{amplsum}) and (\ref{QNNQIM}) 
for $\gamma d\rightarrow pn$ amplitude one obtains:
\begin{eqnarray}
& & \langle p_{\lambda_A},n_{\lambda_B}\mid A\mid \lambda_\gamma,\lambda_D\rangle = 
\sum\limits_{\lambda_2}  {f({\theta_{cm}})\over 3 \sqrt{2s'}}\times \nonumber \\ 
& & 
\ \ \ \ \ \ \left(\langle p_{\lambda_A},{n_{\lambda_B}}|A_{pn}(s,t_n)|
p_{\lambda_\gamma},n_{\lambda_2}\rangle  + \right. \nonumber \\
& & 
\ \ \ \ \ \ \left. \langle p_{\lambda_A},{n_{\lambda_B}}|A_{pn}(s,u_n)|
n_{\lambda_\gamma}p_{\lambda_2}\rangle \right)
\nonumber \\
& & 
\ \ \ \ \ \ 
\int \Psi^{\lambda_D,\lambda_\gamma,\lambda_2}(\alpha_c,p_\perp) {d^2p_\perp\over (2\pi)^2} 
\label{ampl}
\end{eqnarray}
where $t_n = (p_B-{1\over 2}p_D)^2$, $u_n = (p_A-{1\over 2}p_D)^2$  
and $A_{pn}$ is the helicity amplitude of $pn$ scattering, which is factorized from 
the integral. In the  factorization we take into account also the antisymmetry 
of the deuteron wave function  with respect to $p\leftrightarrow n$.
This factorization is justified due to the fact that  at  $\alpha_c={1\over 2}$
the momenta involved in the integration, $p_\perp\le  300$~MeV/c  are much smaller 
than the transferred momenta in the $A_{pn}$ amplitude. For the same reason 
one can approximate the light-cone deuteron wave function that enters in Eq.(\ref{ampl})
through rather well known nonrelativistic deuteron wave function\cite{FS81,gdpn}:
$\Psi^{\lambda_D,\lambda_1\lambda_2} = 
(2\pi)^{3\over 2}\Psi_{NR}^{J_D,\lambda_1,\lambda_2}\sqrt{m}$, where
$\Psi_{NR}^{\lambda_D,\lambda_1,\lambda_2} = 
[u(k) + w(k)\sqrt{{1\over 8}}S_{12}]\xi_1^{\lambda_D,\lambda_1,\lambda_2}$, with
$u(k)$ and $w(k)$ corresponding to the $s-$ and $d-$ waves normalized as
$\int |u(k)|^2 (|w(k)|^2) d^3 k = 1$ and  $\xi_1^{\lambda_D,\lambda_1,\lambda_2}$ represents
the spin component of the wave function.

\section{Predictions for Polarization Observables}

\noindent{\bf A: Definition of Observables}\\
We will discuss several polarization observables of 
reaction (\ref{reaction}) for which there are ongoing  
experimental investigations\cite{Yerevan,RG}. These are:
\begin{itemize}
\item  recoil-proton polarization $P_y$ which corresponds to the measurement of 
asymmetry in the spin component of the protons parallel / antiparallel to the direction of 
$y=\hat q\times \hat p_p$ for the reaction with unpolarized photon and deuteron. 
\item Transfered polarizations $C_{x'}$ and $C_{z'}$, which correspond to the 
measurement of asymmetry in the spin component of the protons parallel / antiparallel to the directions of 
$\hat x' = \hat p_p\times \hat y$ and $\hat p_p$ respectively for the reaction 
with circularly polarized photons and unpolarized deuteron.
\item Cross section asymmetry $\Sigma$ for the reaction with linearly polarized photons
\end{itemize}
These observables are expressed through the helicity amplitudes  
$\langle \lambda_p\lambda_n\mid A\mid \lambda_\gamma,\lambda_d\rangle$ as 
follows\cite{GG,Barannik}:
\begin{eqnarray}
& & f(\theta) P_y = 2 Im \sum\limits_{i=1}^{3}\left[ 
F^\dagger_{i+} F_{[i+3]-} + F_{i-} F^\dagger_{[i+3]+} \right]
\nonumber \\
& & f(\theta) C_{x'} = 2Re \sum\limits_{i=1}^{3}\left[ 
F^\dagger_{i+} F_{[i+3]-}+F_{i-} F^\dagger_{[i+3]+}\right]
\nonumber \\
& & f(\theta) C_{z'} = \sum\limits_{i=1}^{6}\left[\mid F_{i+}\mid^2- \mid F_{i-}\mid^2\right]
\nonumber \\
& & f(\theta) \Sigma = - 2Re \left[\sum\limits_{\pm}
(F^\dagger_{1\pm} F_{3\mp}- F^\dagger_{4\pm} F_{6\mp}) \right.
\nonumber \\
& & \ \ \ \ \ \ 
\left. -F^\dagger_{2+} F_{2-}+ F^\dagger_{5+} F_{5-}\right] 
\nonumber \\
& & f(\theta) = \sum\limits_{i=1}^{6}\sum\limits_{\pm} \mid F_i\pm\mid ^2
\label{observs}
\end{eqnarray}
where $F_{i\pm} = \langle \pm,\pm\mid A\mid 1,2-i\rangle$, for $i=1,2,3$ and 
 $F_{i\pm} = \langle \pm,\mp\mid A\mid 1,5-i\rangle$, for $i=4,5,6$.

\medskip
\medskip

\noindent{\bf B. HRM Predictions}\\
Based on Eq.(\ref{ampl})  one calculates
the  observables defined in Eq.(\ref{observs}) expressing 
them through the helicity amplitudes of $pn$ scattering. 
Derivations are simplified further by using the fact that the momenta relevant in the 
deuteron wave function are $\le 300$~MeV/c. As a result one can restrict by $s$ wave 
contribution in the deuteron wave function only. In this case the radial part of 
the deuteron wave function in Eq.(\ref{observs}) will cancel out and 
one obtains:
\begin{eqnarray}
P_y = - {2Im\left\{\phi^\dagger_5[2(\phi_1+\phi_2) + \phi_3 - \phi_4]\right\}\over
        2|\phi_1|^2 + 2|\phi_2|^2 + |\phi_3|^2+|\phi_4|^2 + 6 |\phi_5|^2}
\nonumber \\
C_{x'} = {2Re\left\{\phi^\dagger_5[2(\phi_1-\phi_2) + \phi_3 + \phi_4]\right\}\over
        2|\phi_1|^2 + 2|\phi_2|^2 + |\phi_3|^2+|\phi_4|^2 + 6 |\phi_5|^2}
\nonumber \\
C_{z'} = {2|\phi_1|^2-2|\phi_2|^2 + |\phi_3|^2 - |\phi_4|^2\over
        2|\phi_1|^2 + 2|\phi_2|^2 + |\phi_3|^2+|\phi_4|^2 + 6 |\phi_5|^2}
\nonumber \\
\Sigma = {2Re\left[|\phi_5|^2- \phi^\dagger_3\phi_4\right]\over
        2|\phi_1|^2 + 2|\phi_2|^2 + |\phi_3|^2+|\phi_4|^2 + 6 |\phi_5|^2},
\label{observs2}
\end{eqnarray}
where off-shell helicity amplitudes of $pn$ scattering are:
\begin{eqnarray}
\phi_1(s,t_n,u_n) & = & \langle +,+\mid A_{pn\rightarrow pn} + A_{pn\rightarrow np}
\mid +,+\rangle \nonumber \\
\phi_2(s,t_n,u_n) & = & \langle +,+\mid A_{pn\rightarrow pn} + A_{pn\rightarrow np}
\mid -,-\rangle \nonumber \\
\phi_3(s,t_n,u_n) & = & \langle +,-\mid A_{pn\rightarrow pn} + A_{pn\rightarrow np}
\mid +,-\rangle \nonumber \\
\phi_4(s,t_n,u_n) & = & \langle +,-\mid A_{pn\rightarrow pn} + A_{pn\rightarrow np}
\mid -,+\rangle \nonumber \\
\phi_5(s,t_n,u_n) & = & \langle +,+\mid A_{pn\rightarrow pn} + A_{pn\rightarrow np}
\mid +,-\rangle.
\label{hamples}
\end{eqnarray}

Due to the relation:
$A_{pn\rightarrow pn/np}  = {A^{I=1}\over 2} +/- {A^{I=0}\over 2}$, in which $I$ is the 
isospin of the $pn$ system one observes that  in the on-shell limit HRM predicts 
a dominance of the isovector channel in $pn$ rescattering at $\theta_{cm}=90^\circ$. 
In this case one has the following features of 
on-shell $\phi$-amplitudes at $\theta_{cm}=90^\circ$:
i) $\phi_5 = 0$ and  ii) $\phi_3 = -\phi_4$.
Furthermore, for any given  isospin state and $\theta_{cm}$  there is a hierarchy 
in helicity amplitudes in the hard regime of the scattering 
(see e.g. \cite{RS,CChM})\footnote{This hierarchy is well founded phenomenologically, 
even with observed finite effects of helicity nonconservation
(see e.g. \cite{CChM}).}:
\begin{equation}
|\phi_{1}|\ge |\phi_{3}|,|\phi_{4}| > |\phi_{5}| > |\phi_{2}|.
\label{hier}
\end{equation}

Based on the above features one can do following rather general 
observations for polarization observables of Eq.(\ref{observs2}): 
\begin{itemize}
\item $P_y$ and $C_{x'}$ should be small at  large $\theta_{cm}$, due to
the fact that on-shell $\phi_5$ approaches zero 
at $\theta_{cm}\rightarrow 90^\circ$. Thus the smallness of $P_y$ and $C_{x'}$ 
at $90^\circ$ will not
necessarily indicate an onset of helicity conserving regime in the 
scattering amplitude. This observation can be checked by looking at 
$\theta_{cm}$ dependence of  $P_y$ and $C_{x'}$. Their increase 
with $\theta_{cm}$ going away from $90^\circ$ will confirm the 
present conjecture\footnote{Inclusion of the $d$ wave in the deuteron wave 
function will not change the result, since the additional terms associated with 
the $d$ wave are proportional to $\phi_5$ too.}.

\item Using relations of Eq.(\ref{hier}), from Eq.(\ref{observs2}) one can 
conclude that the relative sign of $P_y$ and $C_{x'}$ is related predominantly to 
the relative phase of $\phi_5$ and $\phi_1$. 
For example, if real and imaginary parts of both $\phi_5$ and $\phi_1$ have same  
signs then $P_y$ and $C_{x'}$ will have an opposite signs.

\item Based on Eq.(\ref{hier}) on expects $C_{z'}$ to have a positive 
values $\approx$ 0.5 -- 1.

\item The relative sign of $\phi_3$ and $\phi_4$ defines the sign of $\Sigma$.
If isovector channel is dominant in the hard $pn$ rescattering 
then one expects $\Sigma >0$ at $\theta_{cm}=90^\circ$.
\end{itemize}

\vspace{-0.4cm}
\medskip
\medskip

\noindent{\bf C. Numerical Estimates}\\
We discuss the numerical estimates for illustration 
purposes only.  Since there are practically no available data  on 
helicity $pn$ amplitudes  for hard scattering kinematics, we model 
them based on quark-interchange framework of the scattering and the 
fact that HRM predicts the dominance of isovector 
state $NN$ rescattering at $\theta_{cm}=90^\circ$. These two features are 
 reflected in the following parameterization (see e.g.\cite{FGST,BCL,RS}):
\begin{eqnarray}
\phi_1 & = & \phi_1(0)\left[{17\over 62}(F(z_t)+F(z_u)) +  
{14\over 62}(F(-z_t)+F(-z_u))\right]
\nonumber \\
\phi_3 & = & \phi_3(0)\left[{25\over 94}(F(z_t)+F(z_u)) +  
{22\over 94}(F(-z_t)+F(-z_u))\right]
\nonumber \\
\phi_4 & = & \phi_4(0)\left[{ 1\over  4}(F(z_t)+F(z_u)) +  
{ 1\over  4}(F(-z_t)+F(-z_u))\right],
\nonumber \\
\label{1u3u4}
\end{eqnarray}
where $\phi_{i}(0) \equiv \phi^{I=1}_i(0)\approx\phi^{pp}_i(\theta_{cm}=90^\circ)$, ($i=1,2,3,4$) 
and the angular function is defined according to Ref.\cite{RS}:  
$F(z) = 1/[(1+z)(1-z)^3]$, with $z_t= 1+{2t_{n}\over s-4m^2}$, and 
$z_u = -1-{2u_{n}\over s-4m^2}$.
We define $\phi_2$ as: 
\begin{equation}
\phi_2 = {\phi_2(0)\over \phi_1(0)} \phi_1.
\label{phi2}
\end{equation}
Because of (\ref{hier}) the observables of Eq.(\ref{observs2})  
depend weakly on the particular choice of $\phi_2$. 
To asses the values of $\phi_{1,2,3,4}(0)$  we use the phenomenological
parameterizations of \cite{CChM}, which successfully describe the 
available polarization and cross section data on hard $pp$ scattering:
\begin{eqnarray}
& & \phi_{1}(0) = {\phi_{+}+ \phi_{-}\over \sqrt{2}};  \ \ 
\phi_{2}(0) = {\phi_{+} -  \phi_{-}\over \sqrt{2}};  \ \ \phi_{4}(0) = -\phi_3(0)   
\nonumber \\
& & \phi_{\pm,3}(0) = {N\over (s/Gev^2)^4}(B_{\pm,3} + C_{\pm,3}
e^{i[\Psi_{\pm,3}(s) + \delta_{\pm,3}]}), 
\label{phipp}
\end{eqnarray}
where  $\phi_{\pm,3} = a{ln(s/\Lambda^2)\over ln(s/\Lambda_i^2)}$ with  
$\Lambda \equiv \Lambda_{QCD}=0.2$ and all remaining parameters: 
$B_i$, $C_i$, $a$, $\Lambda_i$ are defined in Ref.\cite{CChM} (see Table I).

For  $\phi_5$ we use relation that ensures a vanishing value at 
 $\theta_{cm}=90$ in the on-shell limit\cite{RS}
\begin{equation}
\phi_5 =R_{5-}\phi_1 + R_{5+}(\phi_3 + \phi_4),
\label{phi5}
\end{equation}
where $R_{5\pm}$ is an angular factor defined similar to \cite{RS}:
\begin{equation}
R_{5\pm}(\hat t,\hat u)
= \epsilon\left[{1\over \sqrt{-\hat t}} \pm {1\over \sqrt{-\hat u}}\right].
\label{spflip}
\end{equation}
We consider two values for $\epsilon$: $\epsilon=\sqrt{s-4m^2\over 2}$ corresponding 
to the assumption that the smallness of $\phi_5$ at large angles is related
only to the condition:  $\phi_5(\theta_{cm}=90^\circ)=0$,  
and $\epsilon\approx 0.1$ -- characteristic value obtained from the 
analysis of $\phi_5$ for $pp$ scattering which takes into account an 
additional suppression due to helicity conservation\cite{RS}. 
Note that because of the overall smallness of $\phi_5$ at large $\theta_{cm}$  
the unpolarized cross section is practically insensitive to the particular 
choice of $\epsilon$.

In the hard regime when helicities are conserved $\phi_5$ vanishes and its 
nonzero value is related mainly  to the soft component of $NN$ scattering 
(see e.g. Ref. \cite{BrodTer}). Therefore the fact that one can identify the 
kernel of hard rescattering in Eq.(\ref{ampla}) with the hard kernel of NN scattering 
does not justify the replacement of $\hat t$ and $\hat u$ in  
$R_{5\pm}$ by $t_n$ and $u_n$. 
Furthermore, we will refer such a replacement as an ``on shell'' approximation for $\phi_5$.
Additionally, we consider an  ``off-shell'' approximations in which 
in the first case~(``off-shell I'') we identify  $\hat t = -{s-4m^2\over 2}(1-z_t)$ and 
$\hat u =-{s-4m^2\over 2}(1+z_t)$ and in the second case (``off-shell II'')
$\hat t = -{s-4m^2\over 2}(1+z_u)$ and $\hat u =-{s-4m^2\over 2}(1-z_u)$. 
Note that these are only choices which satisfies the 
condition, $|\hat t|< |\hat u|$ at $\theta_{cm}<90$  (forward angles).
The above ambiguity naturally disappears in the on-shell limit.

Fig.~2 demonstrates the HRM predictions for energy dependences of 
$P_y$, $C_{x'}$ $C_{z'}$ and $\Sigma$ at $\theta_{cm}=90^\circ$. Thick and thin curves 
represent the calculations with parameter $\epsilon$ in Eq.(\ref{spflip}) chosen
$\sqrt{s-4m^2\over 2}$ and $0.1$, respectively. Solid and dashed curves correspond to the 
``on-shell''  and ``off-shell'' approximation for $\phi_5$. Note that at $\theta_{cm}=90^\circ$
both off-shell approximations give an identical results.
According to Eq.(\ref{observs2}) the ``on-shell'' approximation predicts 
$P_y$ and $C_{x'}$  to be exactly zero at $\theta_{cm}=90^\circ$. Thus 
vanishing $P_y$ and $C_{x'}$ do not indicate 
unambiguously the onset of helicity  conservation regime. 
The existing data do not rule out the large values for  
helicity flip amplitudes (thick curves). It is interesting to 
note that within HRM the small value of $C_{z'}$ favors a nonvanishing 
contribution from $\phi_2$ and $\phi_5$. Thus the accurate measurement of 
$C_{z'}$ will have an utmost importance.

\begin{figure}[ht]
\vspace{-0.8cm}
\centerline{\epsfig{height=9cm,width=9cm,file=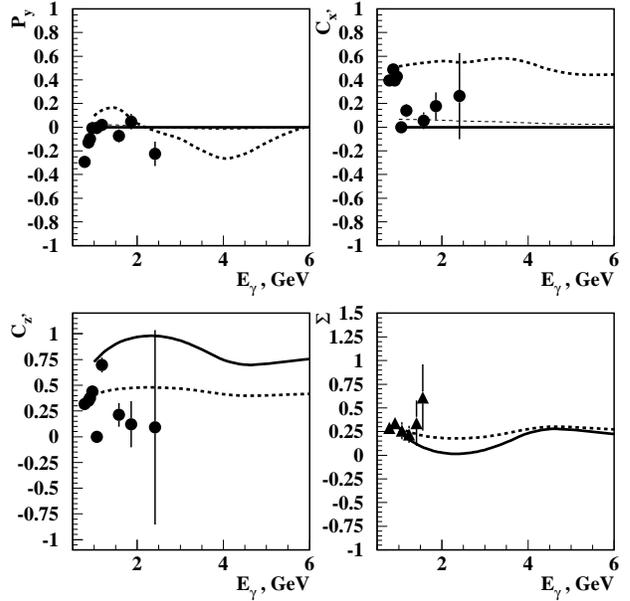}}
\caption{The photon energy dependence of $P_y$, $C_{x'}$ $C_{z'}$ and $\Sigma$ 
at $\theta_{cm}=90^\circ$ photodisintegration of the deuteron.
The curves are described in the text. The $P_y$, $C_{x'}$ and  $C_{z'}$
data are from Ref.\protect\cite{polar}. The  $\Sigma$ data are from 
Ref.\protect\cite{Yerevan}. }
\label{Fig.2}
\end{figure}

The ``on-shell'' and ``off-shell'' approximations can be discriminated unambiguously 
through the study of angular dependences of the observables of Fig.2. 
Fig.~3 demonstrates such a dependence for the reaction with $E_\gamma=4$~GeV. 
The definition of the curves are the same as for Fig.1, with dashed and doted 
curves representing ``off-shell I'' and ``off-shell II'' approximations.
HRM predicts a qualitatively different dependences for  $P_y$, $C_{x'}$ and  $C_{z'}$
for ``on-shell'' and ``off-shell'' approximations of $\phi_5$, 
when no additional suppression due-to helicity conservation is assumed 
($\epsilon={s-4m^2\over 2}$)~(thick curves). 
If the regime of helicity-conservation is established then the difference between 
``on-shell'' and ``off-shell''  approximations become unimportant (thin curves) and 
in both cases HRM predicts a vanishing values for  $P_y$ and $C_{x'}$.  The dominance
of the isovector channel in hard NN rescattering is reflected in the positive 
asymmetry of $\Sigma$.

\section{Summary}
Polarization observables in $\gamma D\rightarrow pn$ have been studied within the hard 
rescattering mechanism of deuteron photodisintegration. 
Within this model  $P_y$, $C_{x'}$ $C_{z'}$ and $\Sigma$ asymmetries are 
expressed through the helicity amplitudes of hard $pn$ scattering.  
At $\theta_{CM}=90^\circ$ HRM predicts a dominance of the isovector channel in the 
hard $pn$  reinteraction.

\begin{figure}[ht]
\vspace{-0.8cm}
\centerline{\epsfig{height=9cm,width=9cm,file=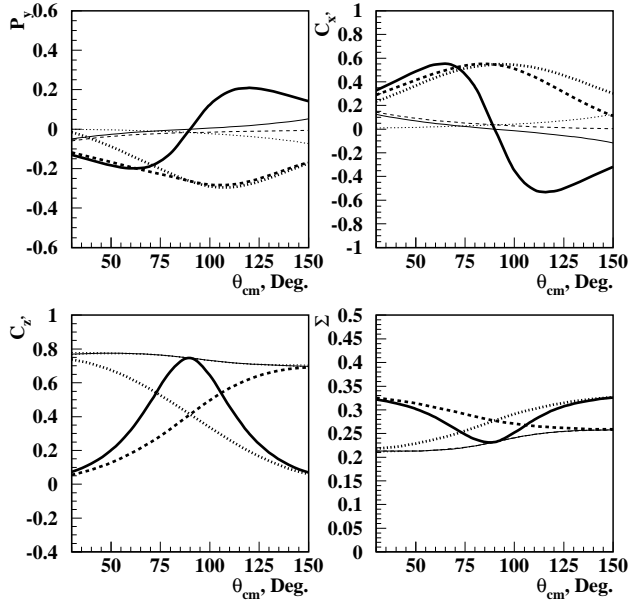}}
\caption{The prediction of $\theta_{cm}$ dependence of 
$P_y$, $C_{x'}$ $C_{z'}$ and $\Sigma$ at $E_{\gamma}=4$~GeV 
photodisintegration of the deuteron. The definition of the curves 
are the same as in Fig.~2.}
\label{Fig.3}
\end{figure}

Based on the general constraints on $NN$ helicity amplitudes we predict 
several qualitative features of $P_y$, $C_{x'}$ $C_{z'}$ and $\Sigma$. These are
the vanishing values of $P_y$, $C_{x'}$ at $\theta_{cm}=90^\circ$ due to 
$\phi^{I=1}_5(\theta=90^\circ)$,  positive large value for $C_{z'}$ if helicity 
conserving regime is established, as well as a  positive sign for 
$\Sigma$. 

Within the quark-interchange framework 
we model the $pn$ helicity amplitudes expressing 
unknown parameters through the existing parameterization of $pp$ amplitudes. 
Our numerical predictions are in reasonable agreement with the existing data, 
indicating that the available data are not sufficient to relate  unambiguously  
the  observed smallness of $P_y$, $C_{x'}$ to the onset  of the  
helicity-conserving regime. Within HRM this smallness can be explained rather 
by the vanishing  $\phi_5$ amplitude for $NN$ scattering at $90^\circ$ in isovector 
channel.
On the other hand the vanishing helicity non-conserving amplitudes within HRM 
predict a sizable asymmetry for $C_{z'}$. Thus it is very important to 
have an  accurate measurement of $C_{z'}$.
In addition, the  study  of the angular dependences of $P_y$, $C_{x'}$ and $C_{z'}$ 
will clarify unambiguously the question  whether the smallness of $P_y$, $C_{x'}$ 
is related to the vanishing $\phi_5$ at $\theta_{cm}=90^\circ$ or the onset of 
helicity conserving regime of high energy scattering. 
The experimental verification of the sign of $\Sigma$ will 
check HRM observation that $\theta_{cm}=90^\circ$ scattering is dominated 
by hard $pn$ rescattering in the isovector channel.

\medskip
\medskip

I would like to thank Drs. S.~Brodsky, L.~Frankfurt, G.~Miller, A.~Radyushkin
and M.~Strikman for many illuminating discussions. I am grateful to Dr. R.~Gilman for 
many useful discussions as well as for providing and explaining the experimental data on 
$P_y$, $C_{x'}$ and $C_{z'}$. Special thanks to Dr. A.~Sirunyan for providing Yerevan 
data on  $\Sigma$. I thank Jefferson Lab for partial support  and 
Institute for Nuclear Theory at the University of Washington for hospitality during the 
completion of this work. This work is supported by U.S. Department of Energy grant 
under contract DE-FG02-01ER41172.


\end{document}